\documentclass[twocolumn, tighten, dvipsnames, floatfix]{aastex63}

\usepackage[]{xcolor}
\usepackage{booktabs}
\usepackage{physics}
\usepackage{longtable}
\usepackage{threeparttablex} % for "ThreePartTable" environment
\usepackage{array}
\usepackage{amsmath}
\usepackage{comment}
\usepackage{changepage}
\usepackage{rotating}
\usepackage{lineno}
\usepackage[flushmargin]{footmisc}
\usepackage{afterpage}
\usepackage{subfigure}

\newcommand{\vect}[1]{\boldsymbol{#1}}

\def\arcsec{\hbox{$^{\hbox{\rlap{\hbox{\lower4pt\hbox{$\,\prime\prime$}}
  }}}$} \ }

\def\arcmin{\hbox{$^{\hbox{\rlap{\hbox{\lower4pt\hbox{$\;\prime$}}
  }\hbox{$\frown$}}}$}}

\usepackage{lipsum}
\usepackage{float}

\shorttitle{Abundance Patterns in the Helmi Stream}
\shortauthors{Limberg et al.}

\begin{document}

\title{Abundance Patterns of $\alpha$ and Neutron-capture Elements in the Helmi Stream} 

%\title{$r$-process Enhancement in a Long-vanished Dwarf Galaxy} 

%The $r$-process Enrichment of the Helmi Stellar Stream, alias S2

\correspondingauthor{Guilherme Limberg}
\email{guilherme.limberg@usp.br}

\author[0000-0002-9269-8287]{Guilherme Limberg}
\affil{Universidade de S\~ao Paulo, Instituto de Astronomia, Geof\'isica e Ci\^encias Atmosf\'ericas, Departamento de Astronomia, \\ SP 05508-090, S\~ao Paulo, Brazil}

\author[0000-0002-7529-1442]{Rafael M. Santucci}
\affiliation{Universidade Federal de Goi\'as, Instituto de Estudos Socioambientais, Planet\'ario, Goi\^ania, GO 74055-140, Brazil}
\affiliation{Universidade Federal de Goi\'as, Campus Samambaia, Instituto de F\'isica, Goi\^ania, GO 74001-970, Brazil}

\author[0000-0001-7479-5756]{Silvia Rossi}
\affil{Universidade de S\~ao Paulo, Instituto de Astronomia, Geof\'isica e Ci\^encias Atmosf\'ericas, Departamento de Astronomia, \\ SP 05508-090, S\~ao Paulo, Brazil}

\author{Anna B. A. Queiroz}
\affil{Leibniz-Institut f$\ddot{u}$r Astrophysik Potsdam (AIP), An der Sternwarte 16, D-14482 Potsdam, Germany}

\author{Cristina Chiappini}
\affil{Leibniz-Institut f$\ddot{u}$r Astrophysik Potsdam (AIP), An der Sternwarte 16, D-14482 Potsdam, Germany}
\affil{Laborat\'orio Interinstitucional de e-Astronomia - LIneA, RJ 20921-400, Rio de Janeiro, Brazil}

\author[0000-0001-8052-969X]{Stefano O. Souza}
\affil{Universidade de S\~ao Paulo, Instituto de Astronomia, Geof\'isica e Ci\^encias Atmosf\'ericas, Departamento de Astronomia, \\ SP 05508-090, S\~ao Paulo, Brazil}

\author[0000-0002-0537-4146]{H\'elio D. Perottoni}
\affil{Universidade de S\~ao Paulo, Instituto de Astronomia, Geof\'isica e Ci\^encias Atmosf\'ericas, Departamento de Astronomia, \\ SP 05508-090, S\~ao Paulo, Brazil}

\author[0000-0002-5974-3998]{Angeles P\'erez-Villegas}
\affil{Instituto de Astronom\'ia, Universidad Nacional Aut\'onoma de M\'exico, Apartado Postal 106, C. P. 22800, Ensenada, B. C., M\'exico}

\author[0000-0002-8262-2246]{Fabr\'icia O. Barbosa}
\affil{Universidade de S\~ao Paulo, Instituto de Astronomia, Geof\'isica e Ci\^encias Atmosf\'ericas, Departamento de Astronomia, \\ SP 05508-090, S\~ao Paulo, Brazil}

\begin{abstract}

%The stellar remnants of dwarf galaxies accreted by the Milky Way can be recognized from their characteristic dynamical signatures. Hence, they provide an avenue to study the early nucleosynthesis processes operating in the environments of these building blocks of the Galaxy. 

We identified 8 additional stars as members of the Helmi stream (HStr) in the combined GALAH+ DR3 and \textit{Gaia} EDR3 catalog. By consistently reevaluating claimed members from the literature, we consolidate a sample of 22 HStr stars with parameters determined from high-resolution spectroscopy and spanning a considerably wider (by $\sim$0.5\,dex) metallicity interval ($-2.5 \lesssim \rm[Fe/H] < -1.0$) than previously reported. Our study focuses on $\alpha$ (Mg and Ca) and neutron-capture (Ba and Eu) elements. We find that the chemistry of HStr is typical of dwarf spheroidal (dSph) galaxies, in good agreement with previous $N$-body simulations of this merging event. Stars of HStr constitute a clear declining sequence in [$\alpha$/Fe] for increasing metallicity up to $\rm[Fe/H] \sim -1.0$. Moreover, stars of HStr show a median value of $+0.5$\,dex for [Eu/Fe] with a small dispersion ($\pm$0.1\,dex). Every star analyzed with $\rm[Fe/H] < -1.2$ belong to the $r$-process enhanced ($\rm[Eu/Fe] > +0.3$ and $\rm[Ba/Eu] < 0.0$) metal-poor category, providing remarkable evidence that, at such low-metallicity regime, stars of HStr experienced enrichment in neutron-capture elements predominantly via $r$-process nucleosynthesis. Finally, the extended metallicity range also suggests an increase in [Ba/Eu] for higher [Fe/H], in conformity with other surviving dwarf satellite galaxies of the Milky Way.

%%For the first time, a significant number of stars of HStr with Eu abundances is presented. 

%the long-vanished dwarf-galaxy progenitor system

% and commensurate with a shared chemical-evolution history.

%In the upcoming years, stars from accreted substructures in the stellar halo will serve as important laboratories to understand the early nucleosynthesis of heavy elements in their long-vanished dwarf-galaxy progenitors that served as building blocks to the Milky Way. 

%This \textit{Letter} constitutes the first systematic investigation of a significant sample of confident member stars of HStr with Eu abundances available. 

\end{abstract}


%\linenumbers

\section{Introduction}
\label{sec:intro}
\setcounter{footnote}{4}

Within the hierarchical assembly paradigm, the Galactic stellar halo (or simply ``halo") is expected to retain the chemodynamical signatures of merging events between the Milky Way and dwarf galaxies of various masses in the past \citep{helmi2008, Helmi2020}. Therefore, studies of surviving satellite galaxies provide insights about the formation of the Milky Way itself. Unfortunately, measurements of elemental abundances in the atmospheres of individual stars from these distant systems are extremely difficult (e.g., \citealt{Tolstoy2009}), hindering our ability to directly investigate their star-forming environments and chemical-evolution histories.

An alternative approach is to pinpoint which stars in the halo were accreted and, out of these, which ones share a common origin. A widely utilized strategy to find the stellar remnants of these ancient building blocks is to search for their clumping in phase space. The first identification of a kinematically-cohesive group of stars with this method was presented in a seminal work by \citet{helmi1999}. Throughout the years, this substructure has been known as the Helmi stream (hereafter ``HStr"; see \citealt{Helmi2020} for a recent review). 

In order to investigate the chemical profile of HStr, \citet{Roederer2010} observed 12 likely members of this substructure. More recently, \citet{Aguado2021} acquired spectra for 7 more candidates. Interestingly, both efforts reported that stars of HStr were enriched in neutron-capture elements predominantly via the ``rapid" process ($r$-process) in comparison to the ``slow" one ($s$-process; see \citealt{Sneden2008} and \citealt{frebel2018} for reviews). In recent studies, \citet{Limberg2021} and \citet{Gudin2021} demonstrated that, indeed, several $r$-process-enhanced (RPE; [Eu/Fe] $> +0.3$ and [Ba/Eu] $< 0.0$) metal-poor ($\rm[Fe/H] < -1.0$; \citealt{beers2005}) stars are dynamically associated with HStr. It appears that chemical-abundance information for larger samples of genuine members of HStr might allow us to investigate the site(s) for the occurrence of the $r$-process in its (now) destroyed progenitor system. In addition, stars of HStr have the enormous advantage of being much closer (and hence brighter) than any surviving dwarf spheroidal (dSph) or ultra-faint dwarf (UFD) satellite galaxy.

%\footnote{Definition of elemental abundance for a star ($\star$) relative to the Sun ($\odot$): [A/B] $= \log (N_{\rm A}/N_{\rm B})_\star - \log (N_{\rm A}/N_{\rm B})_\odot$, where $N_{\rm A}$ ($N_{\rm B}$) is the number density of atoms of element A (B).}

In this Letter, we identified additional stars of HStr in the third data release (DR3) of the Galactic Archaeology with HERMES (GALAH+ DR3; \citealt{GALAHDR3+}) survey combined with \textit{Gaia} Early Data Release 3 (EDR3; \citealt{GaiaEDR3Summary}). We also searched the literature to consolidate a sample covering an [Fe/H] interval considerably wider than previously reported. The extended metallicity range reveals a declining trend in [$\alpha$/Fe] within $-2.0 \lesssim \rm[Fe/H] < -1.0$, providing clear evidence that HStr is the debris of a long-vanished dwarf galaxy. This sample further allowed us to investigate neutron-capture elements. We demonstrate that the [Ba/Eu] pattern is consistent with the dwarf-galaxy progenitor hypothesis. We also confirm that every star of HStr with $\rm[Fe/H] < -1.2$ belongs to the aforementioned RPE class. The employed data is described in Section \ref{sec:data}. Section \ref{sec:anal} contains our dynamical and chemical analyses. The summary of our results and a brief discussion are presented in Section \ref{sec:conc}.

%The employed data is described . The dynamical and chemical analyses are described in Section \ref{sec:anal}. The summary of our conclusions and a brief discussion are 

% considerably  wider  (by∼0.5 dex)  metallicity  interval  (−2.5.[Fe/H]<−1.0)  thanpreviously reported

%We apply a set of kinematic/dynamical criteria to identify additional stars of HStr in the third data release of the Galactic Archaeology with HERMES (GALAH+ DR3; \citealt{GALAHDR3+}) survey combined with \textit{Gaia} Early Data Release 3 (EDR3; \citealt{GaiaEDR3Summary}). Furthermore, we reevaluate the membership of stars previously claimed to be related to this substructure with the same selection. Then, we present the abundance analysis of $\alpha$ (Mg and Ca) and neutron-capture (Ba and Eu) elements of these vetted stars of HStr. Finally, we compare these chemical patterns of HStr with those from dSph and UFD galaxies as well as other accreted halo populations. 

\section{Data}
\label{sec:data}

\subsection{Stream Candidates}
\label{sec:candidates}

\begin{figure}[pt!]
\centering
\includegraphics[width=\columnwidth]{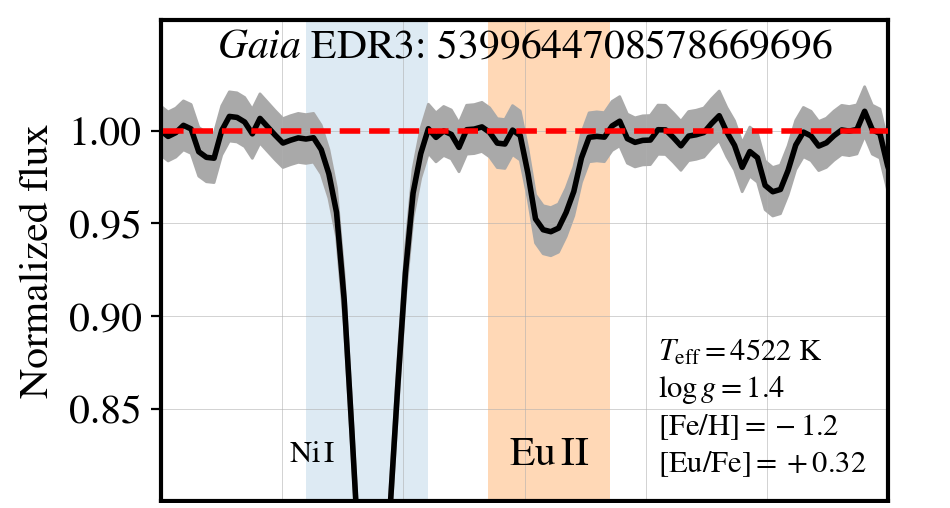}
\includegraphics[width=\columnwidth]{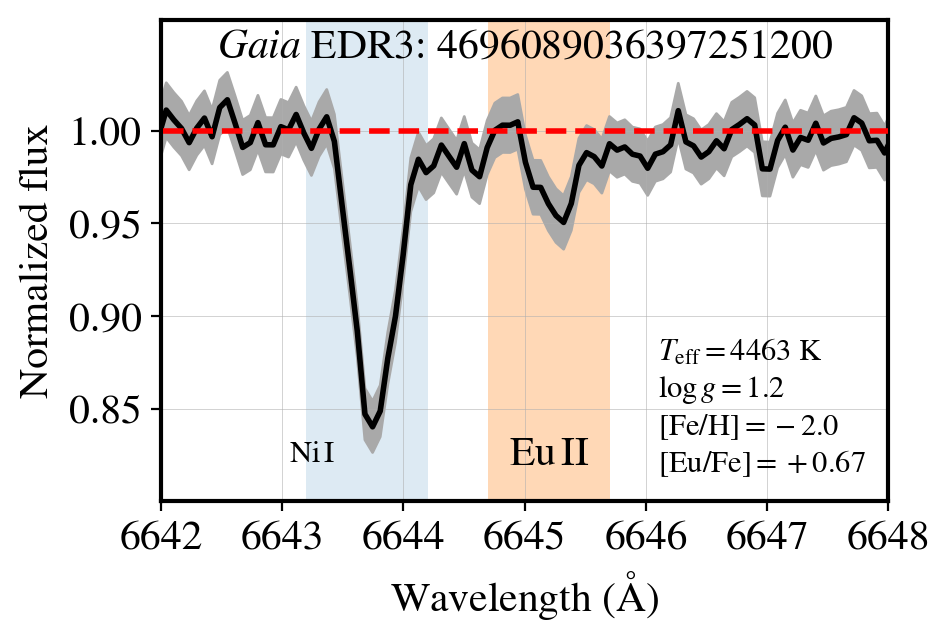}
\caption{{Examples of GALAH+ DR3 spectra in the wavelength region of the \ion{Eu}{2} line (6645\,\AA; orange-shaded area). For the convenience of the reader, we also identify the strong absorption feature as \ion{Ni}{1} (blue stripe). Both panels show stars selected as confident members of HStr (see Section \ref{sec:member}). The black solid lines are the normalized spectra, while the gray-shaded regions represent their respective $1\sigma$ uncertainties. The red dashed line shows the position of the pseudo-continuum to guide the eye.}
\label{fig:spec}}
\end{figure}

We considered the complete GALAH+ DR3 high-resolution ($R \sim$ 28,000) spectroscopic catalog. First, we removed stars with flagged\footnote{Whenever a quality flag is set to ``0" in the GALAH+ DR3 catalog, it represents a reliable and/or real estimate of the given parameter. For details on the quality assessment, we refer the reader to \citet{GALAHDR3+}.} stellar parameters ($\texttt{flag\_sp} \neq 0$) and metallicities ($\texttt{flag\_fe\_h} \neq 0$). Since we are interested in finding stars associated with a substructure of low-metallicity, we further constrain our sample to $\rm[Fe/H] \leq -0.7$ (keeping $\sim$18,000 stars). Moreover, all abundances analyzed (Sections \ref{sec:alpha} and \ref{sec:neutron}) are true measurements ($\texttt{flag\_X\_fe}=0$, where ``X" is any given chemical species). We put particular attention to the estimated [Eu/Fe], since it relies on a single line in GALAH, \ion{Eu}{2} at 6645\,\AA, which is difficult to detect in warm metal-poor stars \citep{Sneden2008}. Hence, all stars with reliable [Eu/Fe] in the sample are in the upper-giant-branch phase ($T_{\rm eff} < 4800$\,K and $\log g < 2.0$). At this wavelength, the typical signal-to-noise ratio is $60 \leq \texttt{snr\_c3\_iraf} \leq 80$ for these objects. We also visually inspected the spectra of stars found to be associated with HStr (Section \ref{sec:member}) to ensure the presence of this absorption feature (see Figure \ref{fig:spec}).

\begin{figure*}[pt!]
\centering
\includegraphics[scale=0.48]{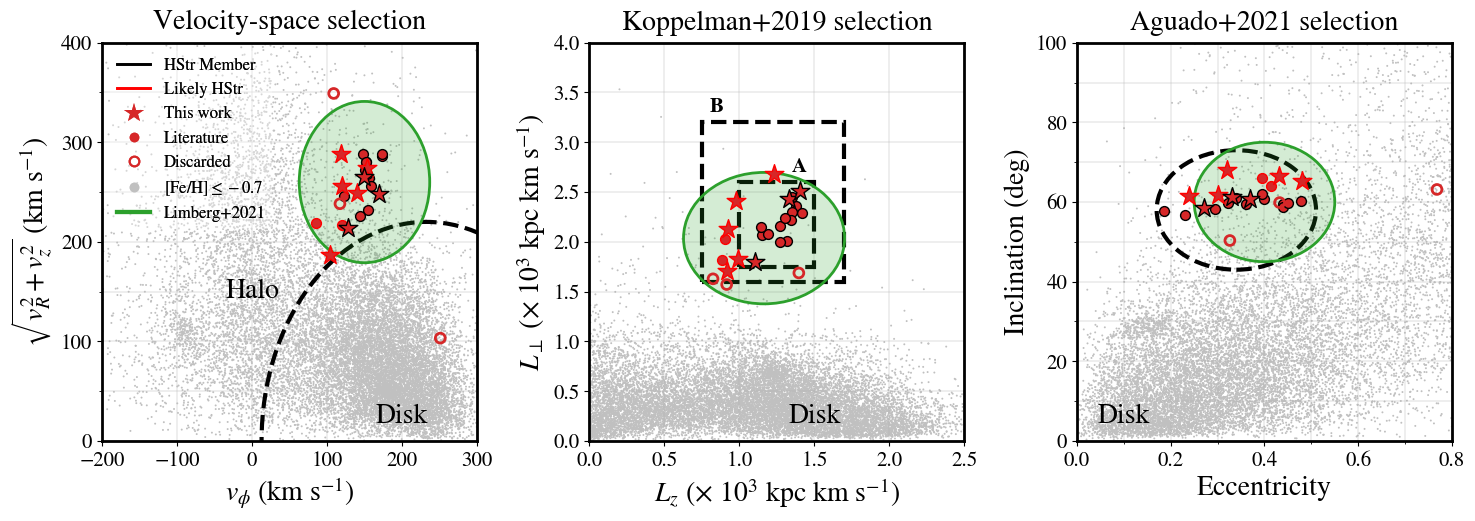}
\caption{Left: velocity distribution defined by $(v_{\phi},\sqrt{v_R^2 + v_z^2})$. Dashed line marks the $||\vect{v}-\vect{v}_{\rm circ}|| \geq 220$\,km\,s$^{-1}$ boundary between disk and halo (see text). Middle: $(L_z,L_{\perp})$. Dashed lines delineate the selection boxes (A and B; see Section \ref{sec:member}) from \citet{koppelmanHelmi}. Right: inclination vs. eccentricity. The dashed ellipse marks the selection from \citet{Aguado2021}. Green ellipses delimit 3-$\sigma$ ranges for the distributions of HStr stars presented by \citet{Limberg2021}. In all panels, the regions dominated by disk-like orbits are also indicated. Stream members are shown as red symbols with black (confident) or red (likely) edges. Open symbols are discarded candidates. Star symbols are members selected from the GALAH sample and circles are from the literature compilation. Gray dots represent low-metallicity ($\rm[Fe/H] \leq -0.7$) stars from GALAH.
\label{fig:member}}
\end{figure*}

To achieve a more complete view of the chemical patterns of HStr, over a large metallicity range, we also compiled stars that had been previously suggested to be associated with the substructure. We require that these stars have atmospheric parameters and abundances estimated from analyses of high-resolution spectra, in conformity with GALAH. The fundamental source is \citet{Aguado2021}, because these authors had already performed a search in the literature for HStr candidates. This sample includes stars observed mostly by the authors themselves and by \citet{Roederer2010}. It also contains stars, attributed to HStr by \citet{yuan2020}, observed over the course of the APOGEE %\footnote{Apache Point Observatory Galactic Evolution Experiment \citep{apogee2017}.}
\citep{apogee2017} survey. %For this work, the information about stars from APOGEE have been updated with DR16 \citep{Jonsson2020}. 
We have further included stars indicated by \citet{Limberg2021} and \citet{Gudin2021} in our literature compilation. Stellar parameters and abundances for these references were obtained during the main effort of the $R$-Process Alliance \citep{hansen2018, ezzedine2020}. \newpage

\subsection{Dynamical Properties}
\label{sec:dyn}

We cross-matched (1.5$''$ search radius) %\footnote{Search radius of 1.5$''$ using the CDS X-Match Service (\url{http://cdsxmatch.u-strasbg.fr/}).} 
all samples with \textit{Gaia} EDR3 \citep{GaiaEDR3Summary} to acquire accurate parallaxes (re-calibrated following \citealt{Lindegren2020_PlxBias}) and absolute proper motions. These astrometric information, the high-resolution spectroscopic data (Section \ref{sec:candidates}), as well as mid- and near-infrared photometry from WISE \citep{WISEsurvey2010} and 2MASS \citep{2MASS}, respectively, were used to estimate %extinctions (in the $V$ band; $\lambda = 542$\,nm) and 
heliocentric distances via isochrone fitting in a Bayesian framework with the \texttt{StarHorse} code \citep{Queiroz2018}. For this exercise, we did not consider targets classified as spurious astrometric solutions ($\texttt{fidelity\_v1} < 0.5$; \citealt{Rybizki2021}). The medians of the resulting posterior probability distribution functions were taken as nominal values (see Appendix \ref{sec:final_list}). We refer to \citet{Queiroz2020} for a complete description of the assumptions regarding stellar-evolution models and priors. We also discarded stars with re-normalized unit weight errors of the reduced astrometric $\chi^2$ outside the recommended interval ($\texttt{RUWE} > 1.4$; \citealt{Lindegren2020_AstromSol}) and those with relative uncertainties $\geq$15\% in their derived distances.

Given the high quality of the spectra at hand, we considered radial velocities (RVs) from the various sources mentioned in Section \ref{sec:candidates}. For the targets observed by \citet{Aguado2021}, neither improved RVs were determined by the authors nor these are provided by \textit{Gaia}'s DRs. Hence, we adopted RVs from the low-resolution spectroscopy of SDSS/SEGUE \citep{sdssYork, yanny2009} and LAMOST \citep{LAMOST1}, with errors no worse than 5\,km\,s$^{-1}$. 

We calculated the orbits of all stars %(GALAH sample and literature compilation) 
for 5\,Gyr forward, with the publicly available library \texttt{AGAMA} \citep{agama}, under the axisymmetric Galactic potential model of \citet{mcmillan2017}. The Galactic parameters are, for consistency, from \citet{mcmillan2017} as well. Specifically, the distance from the Sun to the Galactic center is $R_\odot = 8.2$\,kpc, the circular velocity at this position is $v_{\rm circ} = 232.8$\,km\,s$^{-1}$, and the peculiar motion of the Sun is $(U,V,W)_\odot = (11.10, 12.24,7.25)$\,km\,s$^{-1}$ \citep{schon2010}. We accounted for the uncertainties in distances, proper motions, and RVs, assuming Gaussian profiles for those, by performing 1000 realizations of each star's orbit with a Monte Carlo procedure. The medians of the resulting distributions were taken as our nominal values in the derived dynamical quantities.

\section{Analysis}
\label{sec:anal}

\subsection{Stream Membership}
\label{sec:member}

In order to accomplish a consistent analysis of abundance information (Sections \ref{sec:alpha} and \ref{sec:neutron}), it is necessary to delineate a clear stream membership criteria. Then, we need to implement such selection consistently for all samples described in Section \ref{sec:data}. Conveniently, previous studies carried out similar exercises for HStr.

First, we removed stars with disk-like kinematics utilizing the plane defined by ($v_{\phi},\sqrt{v_R^2 + v_z^2}$), where ($v_R,v_{\phi},v_z$) is the velocity vector in the cylindrical coordinate system. %(radial, azimuthal, and vertical directions, respectively).
Stars with $v_{\phi} > 0$ are in prograde motion, rotating in the same orientation as the disk. For this purpose, we applied the cut $||\vect{v}-\vect{v}_{\rm circ}|| \geq 220$\,km\,s$^{-1}$ (dashed line in the left panel of Figure \ref{fig:member}), where $\vect{v}$ is the complete velocity vector of a given star.

\begin{figure*}[t!]
\centering
\includegraphics[scale=0.48]{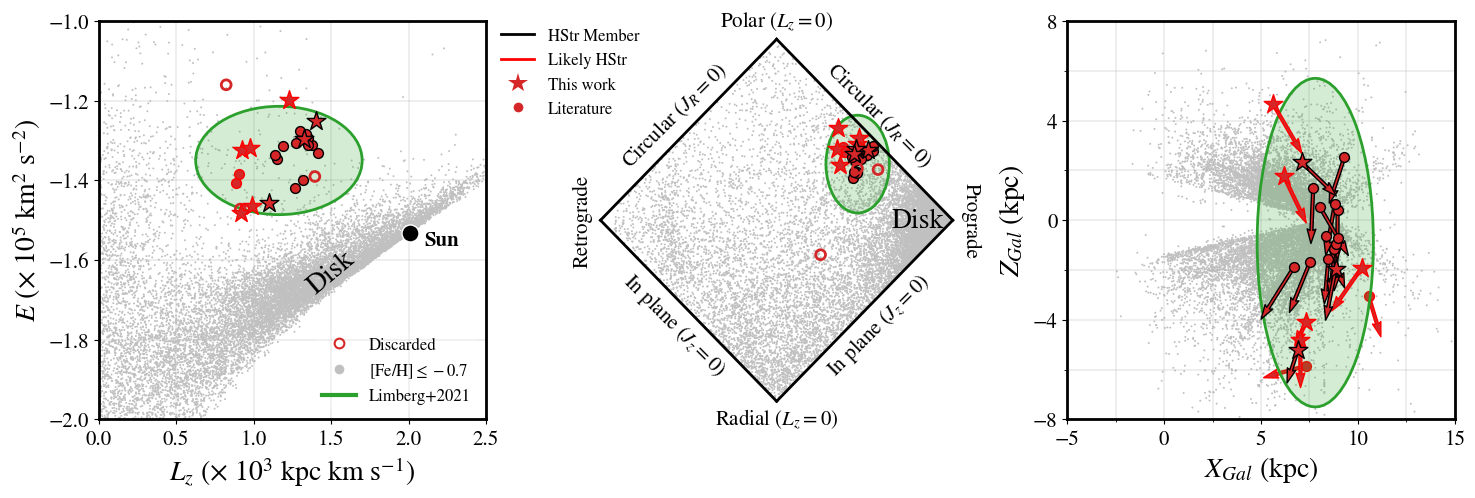}
\caption{Left: $(L_z,E)$. Middle: action-space map. %For comparison, the characteristic locus of other major accreted halo substructures are shown (according to \citealt{myeongSequoia}; see also \citealt{Monty2020} and Section \ref{sec:neutron}) as the orange (Sausage/Enceladus) and blue (Sequoia) rectangles. 
In both panels, the regions dominated by disk-like orbits are also indicated. Right: HStr stars in Cartesian Galactic positions $(X,Z)_{\rm Gal}$. The arrows point in the direction of motion of the stars, scaled by their amplitudes. Green ellipses delimit 3-$\sigma$ ranges for the distributions of HStr stars presented by \citet{Limberg2021}. In all panels, stream members are shown as red symbols with black (confident) or red (likely) edges. Open symbols are discarded candidates. Star symbols are members selected from the GALAH sample and circles are from the literature compilation (Section \ref{sec:candidates}). Gray dots represent low-metallicity ($\rm[Fe/H] \leq -0.7$) stars from GALAH.
\label{fig:tudo}}
\end{figure*}

Second, we employed the criteria suggested by \citet{koppelmanHelmi}, who carried out an in-depth chemodynamical characterization of HStr. These authors' selection boxes are drawn in the middle panel of Figure \ref{fig:member} within the space defined by the in-plane ($L_{\perp}$) and vertical ($L_z$) components of the total angular momentum ($L = \sqrt{L_{\perp}^2 + L_z^2}$). For prograde motion, $L_z > 0$. Although $L_{\perp}$ is not fully conserved in an axisymmetric potential, it is commonly used for identification of substructures since it preserves a reasonable amount of clumping over time (e.g., \citealt{helmi2008, Helmi2020}). Stars that respect the more restrictive box (A) are considered confident members of the stream (symbols with black edges), while objects occupying the more permissive one (B) are taken to be likely associated (red edges). 

%it evolves preserving a certain degree of coherenc

The third criterion used in this work is equivalent to the one described by \citet{Aguado2021}, characterized by orbital inclination ($i = \arccos{(L_z/L)}$; from this definition, $i < 90^{\circ}$ for prograde motion) and eccentricity ($e$). These authors vetted their stream candidates with a 4-$\sigma$ range around the center of the distribution of HStr members originally found by \citet{myeongStreamsAndClumps} %, but updated by \citet{OHare2020}, 
in this parameter space. In the right panel of Figure \ref{fig:member}, the dashed ellipse %, centered at $e=0.34$ and $i=58^{\circ}$, 
reproduces these authors' selection.

%We have defined a box ($40^{\circ} < i < 70^{\circ}$ and $0.15 < e < 0.50$; dashed rectangle in the right panel of Figure \ref{fig:member}) that completely encapsulates their resulting selection, which, as a consequence, is slightly more permissive. 

A comparison between these criteria and the 3-$\sigma$ distributions of HStr stars from \citet{Limberg2021} is also provided in all panels of Figure \ref{fig:member} (green ellipses). The application of this three-step selection to the low-metallicity sample from GALAH yields a total of 8 HStr candidates (red star symbols). Out of these, 3 stars are classified as confident and 5 as probable members. We also evaluated the literature compilation with the same approach, resulting in a total of 14 additional stars to be considered for analysis of chemical-abundance data (red circles). Amongst these stars, 4 are originally from \citet{Roederer2010}, 3 are from \citet{yuan2020}, and 5 were observed by \citet{Aguado2021}. The single star selected from \citet{Limberg2021} has been confirmed to be associated with the stream, while only 1 from \citet{Gudin2021} can be considered a member. The final list of 22 vetted HStr stars, alongside relevant information about them, is provided in Appendix \ref{sec:final_list}.

As a sanity check, we have examined the newly-identified stream members in orbital energy ($E$) and actions ($J_R$, $J_{\phi} = L_z$, and $J_z$). For consistency, these are also compared to the distributions independently found by \citet{Limberg2021}. In Figure \ref{fig:tudo}, we show the 3-$\sigma$ ranges for HStr found by these authors in $(L_z,E)$ and in the action-space map (left and middle panels, respectively). All confident members overlap with the delineated regions, while a couple of the probable ones fall outside of them, but are located near the boundaries. This qualitative inspection helps consolidate our set of criteria as truly representative of HStr.

Despite carrying out a selection exclusively with a kinematic/dynamical approach, stars of HStr are cohesively distributed in configuration space (right panel of Figure \ref{fig:tudo}). These objects are piercing through the Galactic plane, streaming downwards in the Cartesian Galactic position plane $(X,Z)_{\rm Gal}$. Such behavior is commonly interpreted as the partial phase mixing of the debris of a shredded dwarf galaxy \citep{helmi2008,myeongStreamsAndClumps, koppelmanHelmi}.

\subsection{\texorpdfstring{$\alpha$}{} Elements}
\label{sec:alpha}

With our collection of 22 stream members (15 confident and 7 likely), we investigate abundance trends of $\alpha$ elements (Mg and Ca) derived exclusively from high-resolution spectroscopy. In this context, we can compare our [$\alpha$/Fe]--[Fe/H] distributions with those presented by \citet{Aguado2021}, who analyzed a mixture between low- and high-resolution data. We can also compare these chemical-abundance patterns with those from surviving dwarf satellite galaxies of the Milky Way and speculate about the nature of HStr's parent system.

We present the $\alpha$-element profiles in the top row of Figure \ref{fig:abund}. The most prominent, immediately perceptible feature is the clear decrease in [$\alpha$/Fe] (both Mg and Ca) with increasing metallicity, but plateauing at $\sim$0.35 for $\rm[Fe/H] \lesssim -2.0$, characterizing a ``knee" (e.g., \citealt{Matteucci1986}). This point delimits the transition between core-collapse and type Ia supernovae dominated epochs in chemical evolution and was first conjectured for external galaxies by \citet{Matteucci1990}, later observed by \citet{Shetrone2003}. A similar $\alpha$-pattern signature was presented by \citet{Aguado2021}. However, their results were limited to lower metallicities ($\rm[Fe/H] \lesssim -1.5$). Hence, the inclusion of the GALAH sample confirms the continuity of this $\alpha$-element trend of HStr stars up to $\rm[Fe/H] \sim -1.0$ with an excellent agreement. We stress that considering exclusively the newly found members from GALAH, this pattern is still noticeable. Consequently, even if small biases between sources might exist, these qualitative conclusions are robust against them.

The comparison to dwarf satellite galaxies of the Milky Way is presented in Figure \ref{fig:abund} as well. Abundances have been compiled from the SAGA database \citep{SAGA_1, SAGA_4}, favouring references with neutron-capture elements available. The Sculptor dSph \citep{Hill2019} follows a similar sequence in [$\alpha$/Fe] for the same metallicity range ($-2.5 \lesssim \rm[Fe/H] \lesssim -1.0$) to that of HStr stars, but with a steeper decline. With the aid of chemical-evolution models, such feature has been interpreted by \citet{Aguado2021} as a dSph-galaxy progenitor with a slower star-formation rate, but accompanied by a smaller wind efficiency, when compared to Sculptor. This apparent dSph-like origin is in keeping with previous $N$-body simulations of this merging event by \citet{koppelmanHelmi}.

\subsection{Neutron-capture Elements}
\label{sec:neutron}

In this work, we are also interested in evaluating the hypothesis that the progenitor of HStr was enriched in neutron-capture elements (Ba and Eu) predominantly via the $r$-process, as suggested by \citet[and followed-up by \citealt{Aguado2021}]{Roederer2010} and reinforced with a dynamical counterpart to the argument by \citet[also \citealt{Gudin2021}]{Limberg2021}. Here, we consider the abundances of Ba and Eu, since these chemical species serve as diagnostic for $r$-process enhancement in metal-poor stars as first noted by \citet{Spite1992}, although the existence of this kind of objects was established much earlier by \citet{SpiteSpite1978}. \citet{Roederer2010} were capable of determining abundances of these elements for some of their stream candidates. Out of 4 stars observed by these authors and vetted as confident members (Section \ref{sec:member}), 3 of them have [Ba/Fe] and [Eu/Fe] measured. All stream stars from the GALAH sample have non-flagged ($\texttt{flag\_Ba\_fe} = 0$) [Ba/Fe], but only 3 of them have usable ($\texttt{flag\_Eu\_fe} = 0$) Eu abundances. In total, 8 HStr stars can be employed for our investigation of neutron-capture nucleosynthesis in this substructure.

\begin{figure*}[ht!]
\centering
\includegraphics[scale=0.60]{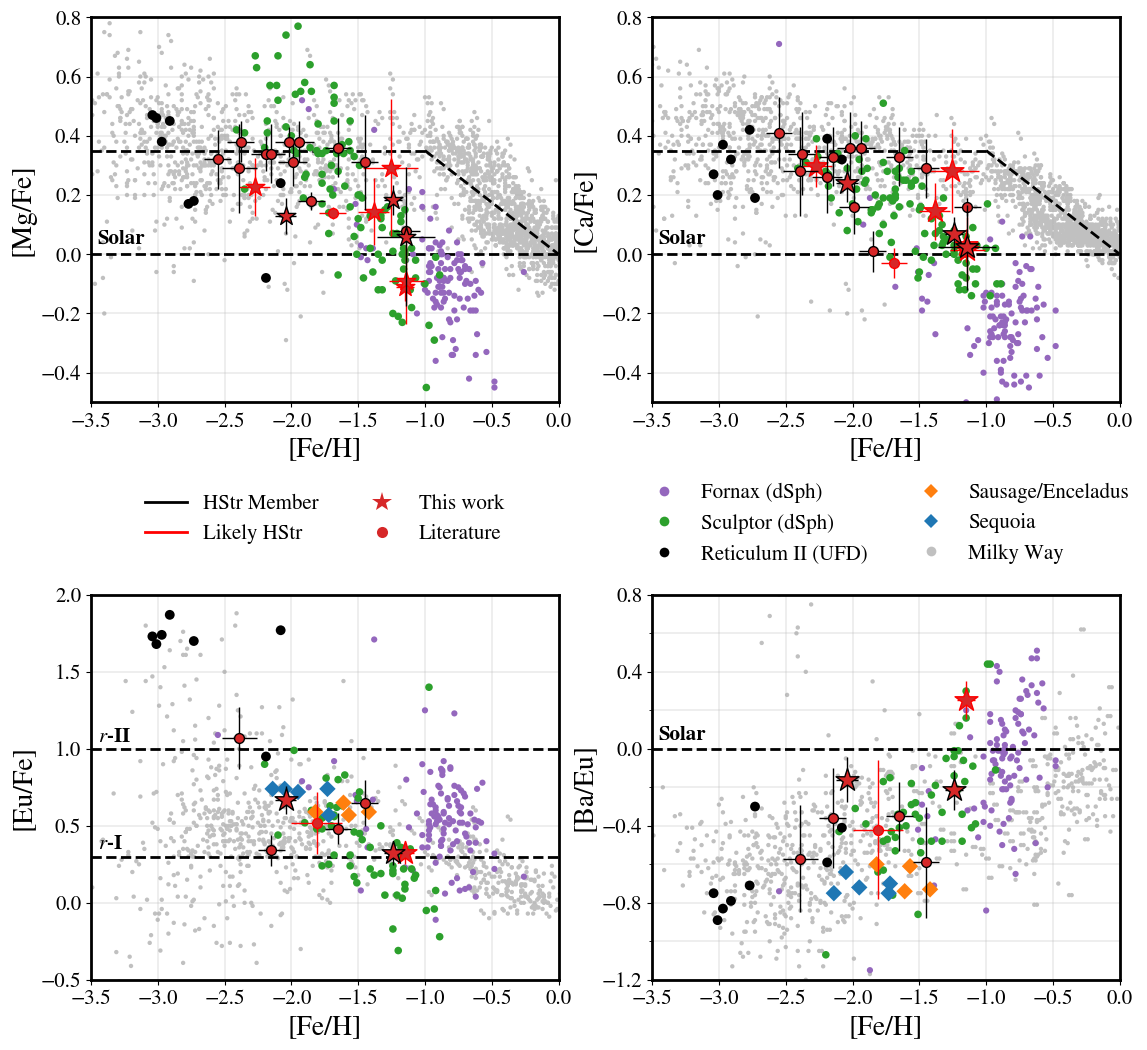}
\caption{Top row: $\alpha$ elements, [Mg/Fe] (left) and [Ca/Fe] (right) versus [Fe/H]. Upper dashed lines approximately follow the evolution of the Milky Way, with a ``knee" at $\rm[Fe/H] = -1.0$ (e.g., \citealt{Matteucci1986}). The solar level is also indicated in both panels. Bottom row: neutron-capture elements. Left: [Eu/Fe]--[Fe/H]. The dashed lines at $\rm[Eu/Fe] = +0.3$ and $+1.0$ delineate the characteristic $r$-process-enhancement of the $r$-I and $r$-II, respectively, classes of RPE metal-poor stars \citep{beers2005}. Right: [Ba/Eu]--[Fe/H]. The solar level is marked as well. In all panels, stream members are shown as red symbols with black (confident) or red (likely) edges. Star symbols are members selected from the GALAH sample and circles are from the literature compilation (Section \ref{sec:candidates}). Colored dots illustrate stars from dSph or UFD galaxies compiled in the SAGA database \citep{SAGA_1, SAGA_4}. Purple: Fornax (dSph; \citealt{Letarte2010, Lemasle2014}). Green: Sculptor (dSph; \citealt{Hill2019}). Black: Reticulum II (UFD; \citealt{Ji2016b}). Dark-blue and dark-red diamonds are stars from Sausage/Enceladus and Sequoia, respectively, observed by \citet{Aguado2021SausageSequoia}. Gray dots in all panels are Milky Way stars, also from SAGA database. We complement these with data from the $R$-Process Alliance \citep{Gudin2021}.
\label{fig:abund}}
\end{figure*}

%PARÁGRAFO ANTIGO
%The [Eu/Fe] as a function of metallicity is displayed in the bottom left panel of Figure \ref{fig:abund}. The majority (7/8) of the HStr members analyzed have $+0.3 < \rm[Eu/Fe] < +0.7$, with a median value of $+0.52$ and median absolute deviation of $\pm0.13$. Despite conducting a selection exclusively from kinematics/dynamics, this is consistent with the other dSph galaxies in SAGA database within the same metallicity range, but the dispersion is even narrower. This apparent dSph-like origin is in remarkable agreement with previous $N$-body simulations of this merging event by \citet{koppelmanHelmi}. We note that the most metal-poor star in the sample, despite showing higher levels of Eu enrichment ($\rm[Eu/Fe] = +1.07$), has [Eu/H] similar to other low-metallicity ($\rm[Fe/H] < -1.2$) stars of HStr. More observations of very metal-poor ($\rm[Fe/H] < -2.0$) stars from this substructure and, of course, measurements of Eu abundances will certainly help constrain the conditions that existed in the early chemical-enrichment site(s) of its progenitor system.

%. While it is correct to say that the [Eu/Fe] is consistent with dwarf galaxies, it is also true that the [Eu/Fe] ratios are very typical of the Milky Way halo. While this may not be a surprise, given that we believe that the halo was formed from accreted debris from smaller galaxies

The [Eu/Fe] as a function of metallicity is displayed in the bottom left panel of Figure \ref{fig:abund}. Every analyzed member of HStr have $\rm[Eu/Fe] > +0.3$, with a median value of $+0.5$ and median absolute deviation of $\pm0.1$. This is, indeed, consistent with the other dSph galaxies in SAGA database within the same metallicity range. However, field metal-poor stars are known for having $\langle \rm[Eu/Fe] \rangle \sim +0.4$ \citep{Shetrone1996, Fulbright2000, Sneden2008}, which is unsurprising given the understanding that the Galactic halo was formed from the accretion of many small dwarf galaxies. Therefore, the elevated Eu abundance alone is not necessarily indicative of the same progenitor, unlike the peculiar $\rm[\alpha/Fe]$--[Fe/H] tendency discussed in Section \ref{sec:alpha}. The most metal-poor star in the sample, despite showing higher levels of Eu enrichment ($\rm[Eu/Fe] = +1.1$), has [Eu/H] similar to other low-metallicity ($\rm[Fe/H] < -1.2$) stars of HStr. Nevertheless, more observations of very metal-poor ($\rm[Fe/H] < -2.0$) stars from this substructure and, of course, detection of Eu will certainly help constrain the conditions that existed in the early chemical-enrichment site(s) of its parent system.

%Fortunately, there now exists a plentiful supply of vetted bright ($V \leq 14$) VMP stars which can be employed for dynamical tagging (with techniques similar to presented in \citealt{yuan2020} or \citealt{Limberg2021}) to select further candidates of HStr.

We can investigate the predominance between the $s$- and the $r$-processes in the neutron-capture nucleosynthesis through [Ba/Eu]. Stars with $\rm[Ba/Eu] > 0.0$ were enriched mostly via the $s$-process, while the birth environments of those with $\rm[Ba/Eu] < 0.0$ were polluted primarily by the $r$-process (\citealt{Spite1992}; see \citealt{Sneden2008} and \citealt{frebel2018} for reviews). In the bottom right panel of Figure \ref{fig:abund}, we present the [Ba/Eu]--[Fe/H] plane. All stars of HStr with $\rm[Fe/H] < -1.2$ occupy the sub-solar region of this parameter space. Therefore, these are moderately RPE ($+0.3 < \rm[Eu/Fe] \leq +1.0$) metal-poor stars, attributed to the so-called $r$-I regime \citep{beers2005}. The only exception is the lowest-metallicity one that belongs to the highly RPE category ($r$-II; $\rm[Eu/Fe] > +1.0$). 

Another important discovery is that stars of HStr follow a similar sequence in [Ba/Eu] as a function of [Fe/H] to that of Sculptor (also perceptible from Figure \ref{fig:abund}), increasing towards higher metallicities. The single star with $\rm[Ba/Eu] > 0.0$ is the most metal-rich in the sample, but is also consistent with Sculptor for this [Fe/H].Accounting for possible systematic effects could shift the [Ba/Eu] down by up to $\sim$0.4\,dex by, for instance, considering non-local thermodynamic equilibrium in abundance calculations \citep{Mashonkina2014}. Despite that, the result that most stars from HStr experienced strong chemical enrichment via the $r$-process would not be altered. Furthermore, the [Ba/Eu] of HStr members would still be in good agreement with stars from Sculptor. This finding corroborates the hypothesis that this surviving satellite of the Milky Way experienced an evolution similar to the progenitor of the stream. If Sculptor is a ``textbook" dSph galaxy \citep{Hill2019}, HStr might be the remnant of an ancient system of similar kind. Hence, spectroscopic studies of its member stars provide a way to refine our understanding about dSph galaxies in general. Most importantly, it represents a marvelous opportunity to study the emergence of $r$-process elements in these environments, especially at the lowest metallicities, but with stars bright enough to have high-resolution spectra readily acquired from ground-based facilities. In the context of the recent observations of gravitational waves \citep{Abbott2017_A}, alongside an electromagnetic counterpart \citep{Abbott2017_B}, of a neutron star merger (GW170817), comparison to $r$-process nucleosynthesis frequencies and yields should help elucidate whether or not this source is responsible, and to what degree, for the production of heavy atomic nuclei.

Likewise, we can compare the Ba- and Eu-to-iron ratios in the stream with those from other prominent kinematic/dynamical halo substructures, also suggested to be of accreted origin. In the bottom row of Figure \ref{fig:abund}, we display the [Eu/Fe] and [Ba/Eu] for metal-poor stars from \textit{Gaia}-Sausage/Enceladus \citep{belokurov2018, helmi2018} and Sequoia \citep{myeongSequoia} with neutron-capture abundances recently published by \citet{Aguado2021SausageSequoia}. These targets cover a narrow metallicity range, within $-2.2 < \rm[Fe/H] \lesssim -1.5$, but comparable to the interval for most HStr stars with such elements available. Overall, these stars from both Sausage/Enceladus and Sequoia are, apparently, more $r$-process rich (medians of $+0.6$ and $-0.7$ for [Eu/Fe] and [Ba/Eu], respectively) than those of HStr ($+0.5$ and $-0.4$) for similar [Fe/H] values. However, given the small numbers considered, it is difficult to achieve meaningful conclusions at this time.

During the preparation of this manuscript, \citet{Gull2021} made available Eu abundances for stars claimed to be associated with HStr. However, their target-selection function was based on pre-\textit{Gaia} data and they did not employ a dynamical selection approach, relying solely on kinematics. After reevaluating their stars, only 2 out of their 12 candidates would be classified as members according to our criteria, both of which were already being considered in our study since they were previously recognized by either \citet[CD--36 1052]{Roederer2010} or \citet[HE 0324--0122]{Limberg2021}. As a consequence, the authors were unable to identify an $\alpha$ ``knee" for a similar metallicity range. Also, many of their stars clearly deviate in both [Eu/Fe] and [Ba/Eu], showing no cohesive sequence as functions of [Fe/H] and a large spread.

%The cause of these discrepancies is their target-selection function, which relied on pre-\textit{Gaia} data and, most importantly, knowledge about the dynamical structure of HStr (not accounting for recent efforts such as \citealt{koppelmanHelmi} and references therein).

Other recent works have utilized orbital and/or phase-space criteria to select members of halo substructures  and/or stellar streams for high-resolution spectroscopy (e.g., \citealt{Monty2020}). However, such efforts are still incipient. This work sheds light on the possibility of taking advantage of dynamical information to, for instance, accelerate the discovery of RPE stars (see the discussion by \citet{Limberg2021_Gemini+SOAR}. Moreover, since these objects likely share a common origin, this approach should be more useful than randomly drawing from extensive lists of cool low-metallicity stars \citep{Limberg2021_Gemini+SOAR}.

\section{Summary}
\label{sec:conc}

%In this Letter, we employed astrometric and spectroscopic data from \textit{Gaia} EDR3 and GALAH+ DR3, respectively, to identify 8 new members of HStr, which allows us to cover a considerably wider (by $\sim$0.5\,dex) metallicity interval ($-2.5 \lesssim \rm[Fe/H] < -1.0$) than previously reported by \citet{Aguado2021}. We also reevaluated candidates from the literature to consolidate a sample of 22 stars of this substructure. Thanks to the now-extended metallicity range, our study clearly shows a declining trend in [$\alpha$/Fe] (Mg and Ca) with increasing [Fe/H]. We were also able to confirm that stars of HStr constitute an $\alpha$-element pattern similar to the Sculptor dSph galaxy up to $\rm [Fe/H] \sim -1.0$. This apparent dSph origin is in good agreement with $N$-body simulations \citep{koppelmanHelmi} of this merging event.

In this Letter, we employed astrometric and spectroscopic data from \textit{Gaia} EDR3 and GALAH+ DR3, respectively, to identify 8 new members of HStr, which allows us to cover a considerably wider (by $\sim$0.5\,dex) metallicity interval ($-2.5 \lesssim \rm[Fe/H] < -1.0$) than previously reported by \citet{Aguado2021}. We also reevaluated candidates from the literature to consolidate a sample of 22 stars of this substructure. Thanks to the now-extended metallicity range, our study clearly shows a declining trend in [$\alpha$/Fe] (Mg and Ca) with increasing [Fe/H]. Considering exclusively the newly found members from GALAH, this pattern is still noticeable. Consequently, our qualitative conclusions are valid even in the presence of small biases between abundances extracted from different sources. We were also able to confirm that stars of HStr constitute an $\alpha$-element pattern similar to the Sculptor dSph galaxy up to $\rm [Fe/H] \sim -1.0$. This apparent dSph origin is in good agreement with $N$-body simulations \citep{koppelmanHelmi} of this merging event.

We confirm that, at low metallicities ($\rm[Fe/H] \lesssim -1.2$), the progenitor system of HStr experienced enrichment in neutron-capture elements predominantly via the $r$-process, as first conjectured by \citet{Roederer2010}. All analyzed stars in this metallicity regime are RPE ones, with median values for [Eu/Fe] and [Ba/Eu] of $+0.5$ and $-0.4$, respectively. The behavior of HStr in [Ba/Eu]--[Fe/H] is also coherent with stars from Sculptor. In particular, the extended metallicity range suggests an increase in [Ba/Eu] for higher [Fe/H], reinforcing the hypothesis that the stream originated from the disruption of a dwarf galaxy of similar kind. Finally, stars from Sausage/Enceladus and Sequoia are, apparently, more $r$-process rich than HStr ones. In the upcoming years, stars from these accreted halo substructures will serve as important laboratories to understand the early nucleosynthesis of heavy elements in their long-vanished dwarf-galaxy progenitor systems.

\acknowledgments

We thank the anonymous referee for a careful review of our work and constructive suggestions that helped improve the manuscript. The authors also thank Jo\~ao A. Amarante, Sean Ryan, and Vinicius Placco for their important suggestions to this work. G.L. acknowledges CAPES (PROEX; Proc. 88887.481172/2020-00) for the funding of his Ph.D. R.M.S. acknowledges CNPq (Proc. 436696/2018-5 and 306667/2020-7). S.R. acknowledges support from FAPESP (Proc. 2015/50374-0 and 2014/18100-4), CAPES, and CNPq. C.C. acknowledges partial support from the ChETEC COST Action (CA16117), supported by COST (European Cooperation in Science and Technology). S.O.S. acknowledges the FAPESP Ph.D. fellowship 2018/22044-3. A.P-V. and S.O.S acknowledge the DGAPA-PAPIIT grant IG100319. H.D.P. thanks FAPESP Proc. 2018/21250-9. F.O.B. acknowledges CAPES (PROEX; Proc. 88887.604787/2021-00). This research has been conducted despite the ongoing dismantling of the Brazilian scientific system. 

This work has made use of data from the European Space Agency (ESA) mission {\it Gaia} (\url{https://www.cosmos.esa.int/gaia}), processed by the {\it Gaia} Data Processing and Analysis Consortium (DPAC, \url{https://www.cosmos.esa.int/web/gaia/dpac/consortium}. Funding for the DPAC has been provided by national institutions, in particular the institutions participating in the {\it Gaia} Multilateral Agreement. This research has also made use of the SIMBAD database and the cross-match service provided by CDS, Strasbourg, France. This work made use of the Third Data Release of the GALAH Survey \citep[]{GALAHDR3+}, which is based on data acquired through the Australian Astronomical Observatory (AAT). We acknowledge the traditional owners of the land on which the AAT stands, the Gamilaraay people, and pay our respects to elders past and present. This paper includes data that has been provided by AAO Data Central (\url{datacentral.aao.gov.au}). This publication also makes use of data products from the Two Micron All Sky Survey \citep{2MASS} and the Wide-field Infrared Survey Explorer \citep{WISEsurvey2010}.

\bibliographystyle{aasjournal}

\bibliography{bibliography.bib}

\appendix

\section{Final list of members}
\label{sec:final_list}

Table \ref{tab:members} contains relevant information about our compilation of stars of HStr from both GALAH and the literature, including uncertainties. The universal identifier for these objects is their \textit{Gaia} EDR3 IDs. In the second column, we provide the classification of the stars either as confident (1) or likely (2) members (Section \ref{sec:member}). Positions ($\alpha, \delta$), proper motions ($\mu_{\alpha} \cos{\delta}, \mu_{\delta}$), and parallaxes ($\varpi$; re-calibrated following \citealt{Lindegren2020_PlxBias}) are included as well. The adopted RV values are also listed. Heliocentric distances ($d_{\rm SH}$) %and extinctions ($A_{\rm V}$) in the $V$ band ($\lambda = 542$\,nm) 
estimated with \texttt{StarHorse} are in the ninth column. Lower and upper limits for these quantities represent the 16th and 84th percentiles of their distributions. The stellar atmospheric parameters and elemental abundances determined from the analyses of the high-resolution spectra (Section \ref{sec:candidates}) of the studied stars are provided by the end of the table. 

\renewcommand{\arraystretch}{1.0}
\setlength{\tabcolsep}{0.29em}

\newpage

%\begin{landscape}
\begin{longrotatetable}
\begin{ThreePartTable}
\begin{TableNotes}
\scriptsize
\item[]\textbf{Notes.} 
\item[]$^{\dagger}$ Newly-recognized members in this work.
\end{TableNotes}

\tiny

\begin{longtable}{>{\tiny}r >{\tiny}c >{\tiny}r >{\tiny}r >{\tiny}r >{\tiny}r >{\tiny}c  >{\tiny}r >{\tiny}c >{\tiny}c >{\tiny}c >{\tiny}c >{\tiny}c >{\tiny}c >{\tiny}c >{\tiny}c}

\caption{Consolidated list of Helmi stream members} \label{tab:members} \\

\hline \multicolumn{1}{c}{\tiny Star name} & \tiny{Class} & \multicolumn{1}{c}{\tiny$\alpha$} & \multicolumn{1}{c}{\tiny $\delta$} & \multicolumn{1}{c}{\tiny $\mu_{\alpha} \cos{\delta}$} & \multicolumn{1}{c}{\tiny $\mu_{\delta}$} & \multicolumn{1}{c}{\tiny ${\varpi}$} & \multicolumn{1}{c}{\tiny RV} & \multicolumn{1}{c}{\tiny $d_{\rm SH}$} & %\multicolumn{1}{c}{\tiny $A_{\rm v}$} &
\multicolumn{1}{c}{\tiny $T_{\rm eff}$} & \multicolumn{1}{c}{\tiny $\log g$} & [Fe/H] & [Mg/Fe] & [Ca/Fe] & [Ba/Fe] & [Eu/Fe]    \\

\multicolumn{1}{c}{\tiny (\textit{Gaia} \tiny EDR3)} & & \multicolumn{1}{c}{\tiny(deg)} & \multicolumn{1}{c}{\tiny(deg)} & \multicolumn{1}{c}{\tiny(mas\,yr$^{-1}$)} & \multicolumn{1}{c}{\tiny(mas\,yr$^{-1}$)} &
\multicolumn{1}{c}{\tiny(mas)} &
\multicolumn{1}{c}{\tiny (km\,s$^{-1}$)} & (kpc)  & (K) & (cgs) & & & & & \\

\hline \hline 
\endfirsthead

\multicolumn{16}{c}%
{{\tablename\ \thetable{} -- continued from previous page}} \\
\hline \multicolumn{1}{c}{\tiny Star name} & \tiny{Class} & \multicolumn{1}{c}{\tiny$\alpha$} & \multicolumn{1}{c}{\tiny $\delta$} & \multicolumn{1}{c}{\tiny $\mu_{\alpha} \cos{\delta}$} & \multicolumn{1}{c}{\tiny $\mu_{\delta}$} & \multicolumn{1}{c}{\tiny ${\varpi}$} & \multicolumn{1}{c}{\tiny RV} & \multicolumn{1}{c}{\tiny $d_{\rm SH}$} & %\multicolumn{1}{c}{\tiny $A_{\rm V}$} &
\multicolumn{1}{c}{\tiny $T_{\rm eff}$} & \multicolumn{1}{c}{\tiny $\log g$} & [Fe/H] & [Mg/Fe] & [Ca/Fe] & [Ba/Fe] & [Eu/Fe]    \\
\multicolumn{1}{c}{\tiny (\textit{Gaia} \tiny EDR3)} & & \multicolumn{1}{c}{\tiny(deg)} & \multicolumn{1}{c}{\tiny(deg)} & \multicolumn{1}{c}{\tiny(mas\,yr$^{-1}$)} & \multicolumn{1}{c}{\tiny(mas\,yr$^{-1}$)} &
\multicolumn{1}{c}{\tiny(mas)} &
\multicolumn{1}{c}{\tiny (km\,s$^{-1}$)} & (kpc) & (K) & (cgs) & & & & & \\                  
\hline \hline 
\endhead

\hline 
\insertTableNotes  \\
\multicolumn{16}{>{\scriptsize}l}{{Continued on next page}}
\endfoot

% & 1 & 1 & 1 & 1 & 1 & 1 &  1

\hline 
\insertTableNotes \\ 
\endlastfoot

2336548524982631808 & 2    & 0.5695   & $-$24.8971 & $-$3.635$\pm$0.013      & $-$8.153$\pm$0.012 & 0.178$\pm$0.024 & 78.8$\pm$2.0    & 5.84$^{+0.74}_{-0.54}$   & %$-$0.194$^{+0.189}_{-0.123}$ & 
5020$\pm$100 & 1.73$\pm$0.30 & $-$1.81$\pm$0.19 & $\dots$ & $\dots$ & $+$0.10$\pm$0.10   & $+$0.52$\pm$0.20    \\

2543922018619120000 & 1    & 6.5826   & 0.6263   & 7.200$\pm$0.045 & $-$22.060$\pm$0.035    &  0.574$\pm$0.059  & 168.0$\pm$3.0   & 1.73$^{+0.13}_{-0.17}$   & %$-$0.477$^{+0.181}_{-0.117}$ & 
6305$\pm$102 & 4.04$\pm$0.52 & $-$2.02$\pm$0.10 & $+$0.38$\pm$0.05   & $+$0.36$\pm$0.12   & $-$0.40$\pm$0.10  & $\dots$ \\
2781151719714308352 & 1    & 12.4166  & 15.5549  & 7.534$\pm$0.055 & $-$29.597$\pm$0.034     & 0.644$\pm$0.053  & 114.0$\pm$3.0   & 1.55$^{+0.11}_{-0.13}$   & %0.230$^{+0.171}_{-0.114}$ & 
6315$\pm$102 & 4.06$\pm$0.52 & $-$1.94$\pm$0.11 & $+$0.38$\pm$0.07   & $+$0.36$\pm$0.09   & $-$0.39$\pm$0.11  & $\dots$ \\
2537418854016682368 & 1$^{\dagger}$    & 15.0695  & 1.2378   & 1.247$\pm$0.038 & $-$20.811$\pm$0.031    &  0.469$\pm$0.032  & 187.3$\pm$1.0   & 2.16$^{+0.17}_{-0.21}$   & %0.039$^{+0.082}_{-0.084}$ & 
5299$\pm$189 & 3.52$\pm$0.26 & $-$1.14$\pm$0.22 & $+$0.06$\pm$0.24   & $+$0.02$\pm$0.14   & $-$0.34$\pm$0.20  &  $\dots$ \\
2518385517465704960 & 1    & 31.7239  & 4.5957   & $-$6.045$\pm$0.043      & $-$37.618$\pm$0.032    &  0.828$\pm$0.053  & 181.8$\pm$5.0   & 1.20$^{+0.06}_{-0.06}$   & %$-$0.212$^{+0.228}_{-0.136}$ & 
6143$\pm$105 & 4.63$\pm$0.51 & $-$1.99$\pm$0.11 & $+$0.31$\pm$0.08   & $+$0.16$\pm$0.09   & $-$0.29$\pm$0.10  & $\dots$ \\
4696863298741726976 & 2$^{\dagger}$    & 40.8047  & $-$65.1743 & $-$0.895$\pm$0.015      & 3.278$\pm$0.013  & 0.157$\pm$0.013 & 255.2$\pm$0.7   & 6.18$^{+0.61}_{-0.74}$   & %$-$0.039$^{+0.088}_{-0.093}$ & 
4898$\pm$128 & 2.09$\pm$0.24 & $-$1.14$\pm$0.13 & $-$0.09$\pm$0.14  & $+$0.02$\pm$0.11   & $+$0.40$\pm$0.14   & $\dots$  \\
2497639347957300096 & 2    & 41.3709  & $-$1.0151  & 3.173$\pm$0.019 & $-$9.668$\pm$0.015  & 0.287$\pm$0.027 & 205.6$\pm$0.1   & 3.84$^{+0.12}_{-0.12}$   & %$-$0.056$^{+0.337}_{-0.234}$ & 
4168$\pm$15 & 0.77$\pm$0.10 & $-$1.69$\pm$0.10 & $+$0.14$\pm$0.02   & $-$0.03$\pm$0.05  & $\dots$ & $\dots$  \\
4696089036397251200 & 1$^{\dagger}$    & 41.4447  & $-$65.5653 & $-$2.994$\pm$0.014      & 1.705$\pm$0.012  & 0.150$\pm$0.013 & 143.0$\pm$0.5   & 6.76$^{+0.54}_{-0.74}$   & %0.166$^{+0.085}_{-0.087}$ & 
4463$\pm$95 & 1.19$\pm$0.25 & $-$2.04$\pm$0.08 & $+$0.13$\pm$0.06   & $+$0.24$\pm$0.07   & $+$0.51$\pm$0.08   & $+$0.67$\pm$0.09    \\
5049085217270417152 & 1    & 41.9060  & $-$36.1075 & 3.187$\pm$0.010 & 12.591$\pm$0.014  & 1.380$\pm$0.023 & 304.7$\pm$0.8   & 0.73$^{+0.01}_{-0.01}$   & %0.429$^{+0.192}_{-0.410}$ & 
6070$\pm$200 & 2.30$\pm$0.30 & $-$1.65$\pm$0.10 & $+$0.36$\pm$0.10   & $+$0.33$\pm$0.10   & $+$0.13$\pm$0.15   & $+$0.48$\pm$0.10    \\
3267948604442696448 & 1    & 51.7594  & 1.5422   & $-$19.554$\pm$0.015     & $-$41.520$\pm$0.014  &  0.968$\pm$0.025    & 186.2$\pm$2.0   & 1.04$^{+0.02}_{-0.03}$   & %0.116$^{+0.180}_{-0.119}$ & 
5170$\pm$100 & 2.50$\pm$0.30 & $-$2.39$\pm$0.13 & $+$0.29$\pm$0.15   & $+$0.28$\pm$0.15   & $+$0.50$\pm$0.20   & $+$1.07$\pm$0.20    \\
4668621380510613248 & 2$^{\dagger}$    & 64.7304  & $-$67.3146 & $-$1.803$\pm$0.017      & $-$0.399$\pm$0.015  & 0.136$\pm$0.016 & 148.1$\pm$0.4   & 6.19$^{+0.54}_{-0.63}$   & %$-$0.001$^{+0.080}_{-0.074}$ & 
4738$\pm$91 & 1.79$\pm$0.26 & $-$1.15$\pm$0.07 & $-$0.11$\pm$0.05  & $+$0.03$\pm$0.06   & $+$0.58$\pm$0.06   & $+$0.33$\pm$0.08    \\
2959451922593403904 & 2$^{\dagger}$    & 77.0546  & $-$25.5669 & $-$9.151$\pm$0.012      & $-$13.330$\pm$0.014  &  0.315$\pm$0.017     & 100.4$\pm$0.6   & 3.36$^{+0.25}_{-0.29}$   & %0.093$^{+0.065}_{-0.061}$ & 
4859$\pm$111 & 2.00$\pm$0.23 & $-$2.27$\pm$0.12 & $+$0.23$\pm$0.10   & $+$0.30$\pm$0.07   & $-$0.07$\pm$0.08  & $\dots$ \\
680735037764135808  & 1    & 122.1393 & 24.3130  & $-$49.704$\pm$0.026     & $-$37.350$\pm$0.019    &  1.130$\pm$0.037   & $-$88.1$\pm$5.0   & 0.88$^{+0.02}_{-0.03}$   & %$-$0.192$^{+0.206}_{-0.134}$ & 
6070$\pm$102 & 4.50$\pm$0.51 & $-$2.38$\pm$0.10 & $+$0.38$\pm$0.07   & $+$0.34$\pm$0.14   & $+$0.50$\pm$0.13   & $\dots$ \\
814087862430876160  & 1    & 142.4193 & 41.0978  & $-$39.395$\pm$0.027     & $-$20.315$\pm$0.021     & 1.186$\pm$0.041   & $-$222.0$\pm$3.0  & 0.88$^{+0.02}_{-0.03}$   & %$-$0.197$^{+0.207}_{-0.145}$ & 
5973$\pm$105 & 4.64$\pm$0.51 & $-$2.19$\pm$0.11 & $+$0.34$\pm$0.06   & $+$0.26$\pm$0.12   & $+$0.40$\pm$0.14   & $\dots$ \\
5399644708578669696 & 1$^{\dagger}$    & 170.7452 & $-$34.1081 & $-$11.629$\pm$0.014     & $-$5.05$\pm$0.011  & 0.183$\pm$0.016 & 70.8$\pm$0.4    & 5.23$^{+0.44}_{-0.45}$   & %0.300$^{+0.082}_{-0.084}$ & 
4523$\pm$87 & 1.45$\pm$0.31 & $-$1.24$\pm$0.07 & $+$0.18$\pm$0.05   & $+$0.07$\pm$0.06   & $+$0.11$\pm$0.07   & $+$0.32$\pm$0.08    \\
840616123070439552  & 1    & 177.6522 & 54.1241  & $-$3.861$\pm$0.011      & $-$1.818$\pm$0.012  & 0.306$\pm$0.022 & $-$277.7$\pm$0.1  & 2.85$^{+0.16}_{-0.27}$   & %$-$0.014$^{+0.196}_{-0.129}$ & 
4907$\pm$25 & 2.16$\pm$0.10 & $-$1.14$\pm$0.10 & $+$0.08$\pm$0.02   & $+$0.16$\pm$0.03   & $\dots$ & $\dots$  \\
6153268944831195648 & 2$^{\dagger}$    & 190.2765 & $-$38.6669 & $-$9.576$\pm$0.015      & $-$11.931$\pm$0.012  &   0.256$\pm$0.018    & $-$43.7$\pm$0.6   & 4.13$^{+0.36}_{-0.43}$   & %0.227$^{+0.068}_{-0.076}$ & 
4897$\pm$131 & 2.26$\pm$0.10 & $-$1.38$\pm$0.12 & $+$0.14$\pm$0.11   & $+$0.15$\pm$0.10   & $+$0.04$\pm$0.12   &  $\dots$ \\
3630429997250686592 & 2$^{\dagger}$    & 202.5277 & $-$7.6950  & $-$6.231$\pm$0.025      & $-$6.512$\pm$0.014  & 0.194$\pm$0.023 & $-$228.9$\pm$0.9  & 5.43$^{+0.76}_{-0.59}$   & %0.042$^{+0.123}_{-0.104}$ & 
4962$\pm$193 & 2.43$\pm$0.33 & $-$1.26$\pm$0.20 & $+$0.29$\pm$0.23   & $+$0.28$\pm$0.14   & $+$0.05$\pm$0.20   &  $\dots$ \\
3742101345970116224 & 1    & 205.8611 & 15.5752  & $-$49.997$\pm$0.017     & $-$14.872$\pm$0.011    & 1.898$\pm$0.029    & $-$285.9$\pm$0.8  & 0.54$^{+0.01}_{-0.00}$   & %1.155$^{+0.155}_{-0.060}$ & 
5290$\pm$200 & 1.10$\pm$0.30 & $-$2.15$\pm$0.10 & $+$0.34$\pm$0.10   & $+$0.33$\pm$0.10   & $-$0.02$\pm$0.24  & $+$0.34$\pm$0.10    \\
1275876252107941888 & 1    & 226.7244 & 30.0102  & 13.941$\pm$0.009      & $-$7.761$\pm$0.013  & 0.683$\pm$0.027 & $-$280.3$\pm$0.8  & 1.47$^{+0.06}_{-0.06}$   & %$-$0.444$^{+0.412}_{-0.257}$ & 
4330$\pm$200 & 0.60$\pm$0.30 & $-$1.45$\pm$0.10 & $+$0.31$\pm$0.16   & $+$0.29$\pm$0.10   & $+$0.06$\pm$0.25   & $+$0.65$\pm$0.15    \\
1770226575557939840 & 1    & 322.8380 & 13.1278  & 4.082$\pm$0.012 & $-$16.745$\pm$0.010    &   0.258$\pm$0.022  & 71.4$\pm$0.1    & 4.06$^{+0.34}_{-0.28}$   & %0.079$^{+0.194}_{-0.126}$ & 
4975$\pm$16 & 1.82$\pm$0.10 & $-$1.85$\pm$0.10 & $+$0.18$\pm$0.03   & $+$0.01$\pm$0.07   & $\dots$ & $\dots$ \\
6536758912468029184 & 1    & 351.2973 & $-$39.9912 & 4.400$\pm$0.023 & $-$7.736$\pm$0.028  & 0.561$\pm$0.022 & 286.8$\pm$0.8   & 1.80$^{+0.10}_{-0.12}$   & %$-$0.081$^{+0.175}_{-0.112}$ & 
6650$\pm$200 & 3.70$\pm$0.30 & $-$2.55$\pm$0.10 & $+$0.32$\pm$0.10   & $+$0.41$\pm$0.12   & $-$0.31$\pm$0.30  & $\dots$ 

\end{longtable}
\end{ThreePartTable}
\end{longrotatetable}
%\end{comment}

\setcounter{table}{0}

\end{document}